\documentclass[aps,twocolumn,showpacs,preprintnumbers,pra,10pt]{revtex4-1}
\usepackage{graphicx,dcolumn,bm,amsmath,amssymb,hyperref}
\usepackage[english]{babel}

\begin{document}
\title{Quench dynamics and statistics of measurements for a line of quantum spins in two dimensions}
\author{Jonathan Lux$^{1}$}
\email{lux@thp.uni-koeln.de}
\author{Achim Rosch$^{1}$}
\affiliation{$^1$ Institute for Theoretical Physics, University of Cologne, D-50937 Cologne, Germany}
\date{\today}

\begin{abstract}
Motivated by recent experiments, we investigate the dynamics of a line of spin-down spins embedded in
the ferromagnetic spin-up ground state of a two-dimensional xxz model close to the Ising limit. 
In a situation where the couplings in x and y direction are different, 
the quench dynamics of this system is governed by the interplay of one-dimensional excitations (kinks and holes) moving along the line and single-spin excitations evaporating into the two-dimensional background.
A semiclassical approximation can be used to calculate the dynamics of this complex quantum system.
Recently, it became possible to perform projective quantum measurements on such spin systems,
allowing to determine, e.g.,  the z-component of each individual spin. We predict the statistical properties of 
such measurements which contain much more information than correlation functions.
\end{abstract}

\maketitle
\section{Introduction}

In a projective quantum measurement of a wave function $|\Psi\rangle$, one measures the (non-degenerate) eigenvalue $\alpha$ of some observable $A$ with the probability $|\langle \Phi_\alpha |\Psi \rangle|^2$, where
 $A | \Phi_\alpha\rangle=\alpha | \Phi_\alpha\rangle$. This textbook example of a measurement process is, however, not realized in a typical condensed matter experiments.
When one measures thermodynamic, transport or spectroscopic properties, or performs a pump-probe experiment, one usually
 obtains correlation functions of static or sometimes time-dependent many-particle states.
Therefore, also most theoretical approaches for such systems focus on the computation of expectation values and correlation functions. 

Ultracold atoms provide new opportunities to perform quantum simulations of many-particle systems. In these systems it is possible to perform new
classes of experiments, which go beyond the measurement of simple expectation values or correlation functions \cite{dalibard06,lamacraft07,schmiedmayer08,eckert08}.
A major step forward is the  recent development \cite{ertmer02, greiner08, porto09, greiner09,Weitenberg11} of  {\it quantum microscopes}.
With those, one can measure the position of each single atom in an optical lattice  with fidelities close to 100\% \cite{greiner09,Weitenberg11}.
Therefore, one can realize projective quantum measurements of a many-particle system. 
Repeating such an experiment several times (or by analyzing a single-shot experiment of a large, translationally invariant system) one can in principle obtain all diagonal elements of
 the many-particle density matrix in the position basis.
In combination with other transformations, also non-diagonal elements are accessible. For example, it was possible to measure non-local quantities defining string-order in one dimension\cite{Endres11}.
There are also other experimental setups which analyze the statistics of quantum measurements in many-particle systems, for example by studying the interference of Luttinger liquids \cite{schmiedmayer08}.

Our study has mainly been motivated by a setup realized in Munich \cite{Weitenberg11,Fukuhara13,fukuhara13b}. Using two hyperfine states of $^{87}$Rb atoms forming a Mott insulator in an optical lattice, these
experiments realized a spin-$\frac{1}{2}$ Heisenberg model in two dimension with separately tunable couplings in spatial $x$ and $y$-direction.
Using the quantum microscope, it is possible to write an arbitrary initial spin-configuration.
Then the quantum dynamics described by the Heisenberg model is switched on.
Finally, after a time $t$ a projective quantum measurement is performed, where the $z$-component of each spin in the lattice is measured. The operators $S^z_i$ define a complete set of commuting observables.
By repeating such an experiment for an $N$-spin system one can extract the full quantum-mechanical probability distribution describing the $2^N$ spin configurations. In this way, the $2^N$ diagonal
elements of the density matrix can be obtained. 
In Ref.~\cite{Fukuhara13}, for example, a $1d$ line of down-spins (see Fig. \ref{fig1}a) in a $2d$ background of up-spins has been prepared as the initial state.
In this experiment only the coupling perpendicular to this line was switched on, therefore this setup effectively realized several copies of 
a single down-spin in $1d$. Due to spin-conservation and the one-dimensional dynamics, this setup realizes the quantum mechanical dynamics of a single particle (the down spin) hopping by spin-flip processes in a $1d$ lattice. The experiment was able to reconstruct 
the time evolution of the probability distribution function of this particle. Similarly, for an initial state with two down-spins, one can trace
the physics of a two-magnon bound state \cite{fukuhara13b}.

In this paper we study the same initial configuration, a $1d$ line of down-spins in a $2d$ ferromagnetic spin-up background, see Fig.~\ref{fig1}a.
But in contrast to the experiment in Ref.~\cite{Fukuhara13}, we study the setup with a true two-dimensional anisotropic coupling.
We further modify the experimental setup by investigating a quantum spin model close to the Ising limit.
Experimentally, this limit can be approached by adjusting the trapping laser beams and using a suitably chosen Feshbach resonance \cite{lukin03, bloch_rmp}.
As discussed in more detail below, this setup shows several interesting features:
(i) it is an example of a many-particle quantum problem tractable by semiclassical techniques,
(ii) it is characterized by a zoo of strongly interacting quasiparticles (kinks and holes), exhibiting one-dimensional dynamics,
(iii) due to the possibility to emit spin excitations into the $2d$ background, it is an open quantum system which does not equilibrate but ultimately freezes out,
(iv) the dynamics shows pronounced algebraic long-time tails, 
(v) the final state is characterized by a distribution of peculiar many-particle bound states, approximately given by line segments of down-spins dressed by quantum fluctuations. 
Standard correlation functions are not sufficient to describe the resulting final state. We will therefore calculate directly the distribution functions characterizing these many-particle bound states.

\section{The model}

In the following, we will consider the $2$-dimensional spin $1/2$ ferromagnetic xxz Heisenberg model 
\begin{widetext}
\begin{equation}
\begin{split} \label{ham}
 \hat{H}  = & -J \sum\limits_{i_x, j_y}\left[ \frac{1}{2} \left(\hat{S}^+_{i_x,j_y} \hat{S}^-_{i_x+1,j_y} + \hat{S}^+_{i_x+1,j_y} \hat{S}^-_{i_x,j_y} \right) +\Delta \hat{S}^z_{i_x,j_y} \hat{S}^z_{i_x+1,j_y} \right] \\
 & - J \alpha \sum\limits_{i_x, j_y}\left[ \frac{1}{2} \left(\hat{S}^+_{i_x,j_y} \hat{S}^-_{i_x,j_y+1} + \hat{S}^+_{i_x,j_y+1} \hat{S}^-_{i_x,j_y} \right) +\Delta \hat{S}^z_{i_x,j_y} \hat{S}^z_{i_x,j_y+1} \right] 
\end{split}
\end{equation}
\end{widetext}
We consider the Ising limit $\Delta \gg 1$. The coupling in the $y$-direction, $\alpha J$, is assumed 
to be stronger than the coupling $J$ in the $x$-direction: $ \alpha-1 \gg 1/\Delta$.

 \begin{figure}[htb]
 \includegraphics[width=0.7 \linewidth]{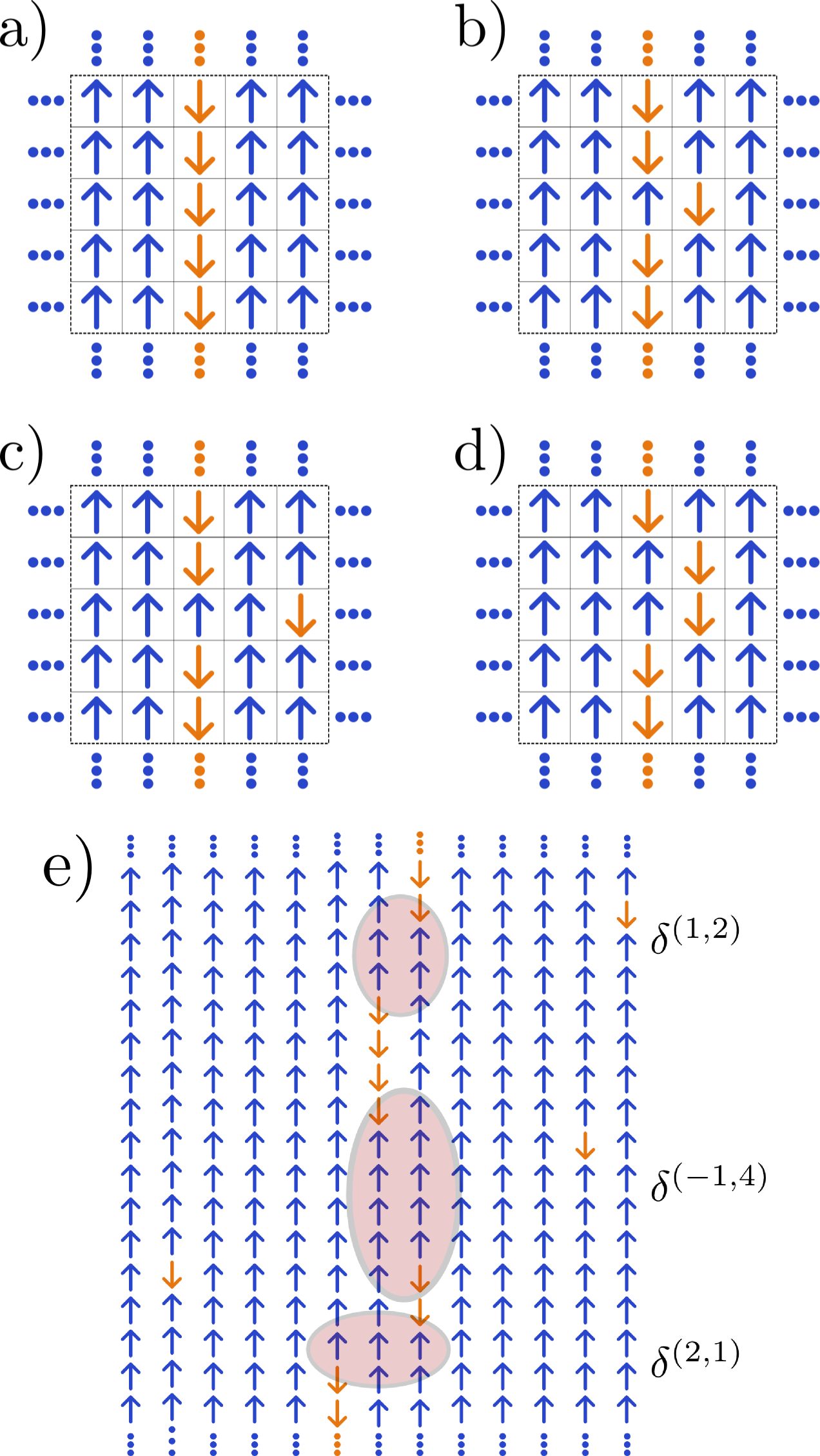}
 \caption{
 a) Initial state: A $1d$ line of down-spins is prepared in a $2d$ background of up-spins.
Such states can be realized with ultracold atoms in optical lattices\cite{Weitenberg11,Fukuhara13}.
b) The only state (up to translations and reflection) that can be reached from the initial state by a single spin flip. Such states arise to leading order in $1/\Delta$ when a quasiparticle description is used, see Eq.~(\ref{SWtrafo}). 
c) A state with two quasiparticles, a hole ($h$) and a free spin ($f$). d) Kink ($k$) and antikink ($\bar k$) excitations. $h$, $k$ and $\bar k$ can only propagate parallel to the line.
 e) Example of more complex defects, $\delta^{(n,m)}$, created by scattering events.  Here $n$ ($m$) describes the shift of the line in $x$ ($y$) direction, respectively.
\label{fig1} }
 \end{figure}
A characteristic property of the model (\ref{ham}) is that for $\Delta \gg 1$ its spectrum is characterized by a
large number of different many-particle bound states and quasi-bound states with extremely long life times. 
Consider, for example, a finite number, say $N_\downarrow$, of down-spins in a ferromagnetic background.
In the Ising limit, $\Delta = \infty$, each `compact' spin configuration (a spin configuration where all down-spins form a connected domain of arbitrary shape) is such a bound state.
For $\infty > \Delta \gg 1$ these bound states get dressed by quantum fluctuations, some of them hybridize with other bound states and some of them can  become unstable.

Consider, for example, a line of down spins oriented along the $y$-direction.
The energy of this configuration relative to the ferromagnetic background for $\Delta \to \infty$ is $\Delta J (N_\downarrow+\alpha)$.
It is generically separated by an energy of order $\Delta J$ from most other spin-configurations.
Splitting up a single down-spin costs, e.g., the energy $\Delta J \alpha$. For large $N_\downarrow$ the system can, however, also gain energy by forming a more compact spin structure. 
When (small) spin-flip terms are considered, a single spin gains kinetic energy of order $J \ll \Delta J$, while
bound states of many spins can gain  by quantum fluctuations only energies of order $J/\Delta$ or smaller.
For $\Delta \gg 1$, low-order spin flips terms will therefore generically {\em not} destroy the many-particle bound states, as they are protected by energies of order $\Delta J$.
For very long lines and commensurate values of $\alpha$ (e.g., $\alpha=2$ and $N_{\uparrow}\ge 13$), complex quantum-tunneling processes exist, where
the energy-gain obtained by forming a compact domain can be used to emit a single down-spin.
In such a situation, the line of down-spins is not a true bound state but only a quasi-bound state.
As the decay channel involves, however, the rearrangement of a large number of spins ($6$ in the present example), it can safely be neglected on all experimentally relevant time scales.

The qualitative discussion given above has shown, that in the considered model a spectrum of many-particle bound states exists.
As in the Munich experiment \cite{Fukuhara13}, we consider the following quench problem:
The initial state $|\Psi(0) \rangle$ is obtained by writing a line of down-spins into the ferromagnetic groundstate $|\, {\rm FM}_{\uparrow}\, \rangle$ (see Fig.~\ref{fig1}a)
\begin{equation}
|\Psi(0) \rangle =\prod_i S_{0,i}^-\, \, |\, {\rm FM}_{\uparrow}\, \rangle \label{psi0}
\end{equation}
The line is written along the $y$-directions where all couplings are stronger by a factor $\alpha$. At time $t=0$, the quantum dynamics according to Eq.~(\ref{ham}) is switched on.
For large $N_\downarrow$, the initial state only has a negligible overlap, $ \sim \exp(-N_\downarrow/(8 \alpha^2 \Delta^2))$ (see below), with the corresponding many-particle (quasi-)bound state as the latter is dressed by quantum fluctuation.
Therefore the initial state will decay by evaporating excitations.
During this evaporation process, however, a large variety of smaller many-particle bound states will form. Characterizing their distribution is a main goal of the following discussion. 

\section{Zoology of excitations and their interactions}\label{zoology}

To compute the quantum dynamics of the system, we use that in the Ising case, $\Delta = \infty$,
the initial state is an exact eigenstate. For $\Delta \gg 1$ it can therefore be described by a low density of quasiparticles.
To identify the quasiparticles and their dynamics to leading order in $1/\Delta$, we use that the many-particle Hilbert space  for large $\Delta\gg 1$ splits into sectors separated by multiples of $J \Delta$ and $\alpha J \Delta$. The spin configurations shown in Fig.~\ref{fig1}b,c and d, for example, have an energy which is
is larger by $2 \alpha \Delta J$ compared to the configuration in Fig.~\ref{fig1}a.
To include the effects of spin-flips, we then perform degenerate perturbation theory.
In a first step, all matrix elements connecting sectors with different energies are ignored, as they are of higher order in $1/\Delta$.
This allows us to identify the quasiparticles relevant for our study: Fig.~\ref{fig1}c can be described by two quasiparticles.
First, a hole (h) is obtained by replacing on the line a down-spin by an up-spin.
This hole can only move {\em along} the line with a hopping rate given by $\alpha J/2$.
Second, a free spin (f) (right side of Fig.~\ref{fig1}c) can propagate in both, the $x$- and $y$-direction, with hopping rates $J/2$ and $\alpha J/2$ respectively.
In Fig.~\ref{fig1}d, there are also two quasiparticles, a kink ($k$) and an antikink ($\bar{k}$).
They are defined by points where the line of down spins takes one step in the negative or positive $x$-direction, respectively.
Similar to the holes, the kinks can propagate only along to the line.
They are more heavy than holes as their hopping rate is given by $J/2$ and therefore smaller.
More generally, one has to consider more complex defects $\delta^{(n,m)}$. These are configurations, where
the line of down spins is shifted by $n$ steps in $x$-direction and $m$ steps in the $y$-direction, see  Fig.~\ref{fig1}e
for some examples. To leading order in $1/\Delta$ only $h=\delta^{(0,1)}$, $k = \delta^{(-1,0)}$, $\bar{k}=\delta^{(-1,0)}$ and $f$ are mobile.
All others defects have hopping rates of order $J/\Delta^j$ with $j \ge 1$. Therefore, they move only on larger timescales.
\begin{figure*}[t]
 \includegraphics[width=0.95\textwidth]{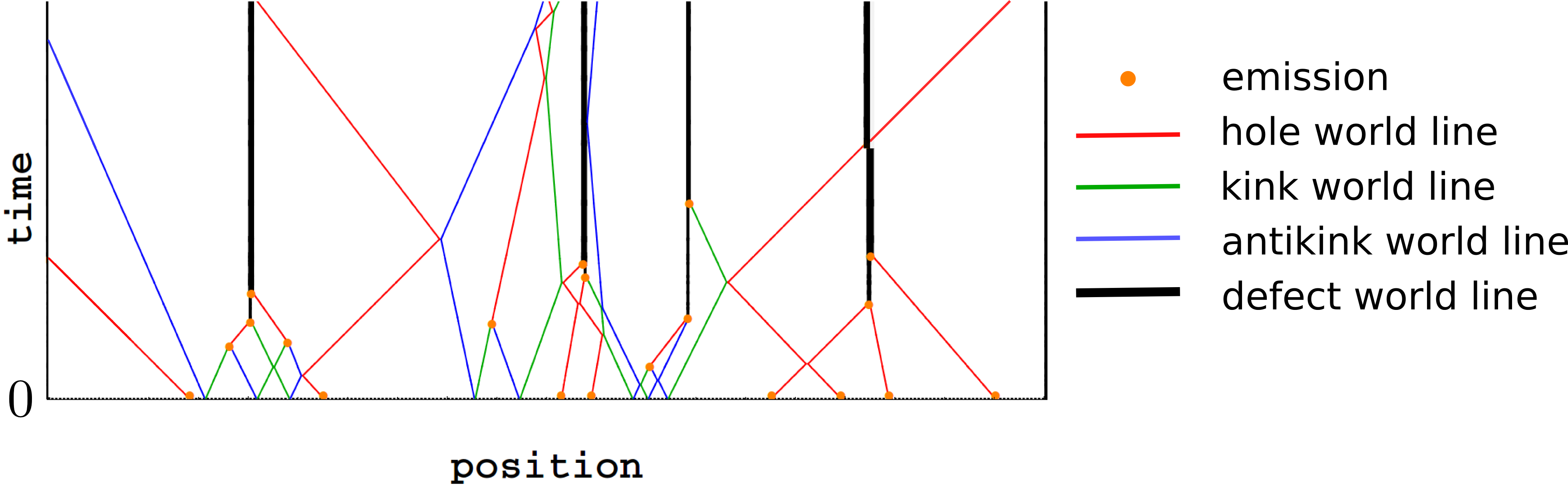}
 \caption{
Typical worldlines describing the semiclassical dynamics. At t=0 an ensemble of kink-antikink pairs (green and blue worldlines) and holes (red) is created.
The creation of a hole is accompanied by the emission of a single spin (only shown as a yellow dot) in the direction perpendicular to the line, see Fig.~\ref{fig1}.
Subsequent scattering can lead to further emission of single spins and the formation of larger defects (black), which are immobile to order $1/\Delta$. 
\label{fig2} }
 \end{figure*}

While the dynamics is governed by terms of the order $\Delta^0$, one has to consider effects of order $1/\Delta$ to calculate
the initial wavefunction $|\Psi(t=0)\rangle$ in terms of the quasiparticles.
To derive a quasiparticle picture one recalls that quasiparticles are defined as (approximate) eigenstates
of the Hamiltonian living within a given energy sector. We therefore used a Schrieffer-Wolff transformation
to derive an effective quasiparticle description. 
In a Schrieffer-Wolff transformation, one constructs a unitary transformation $U=e^{S}$ with $S=-S^\dagger$, such that in the transformed (quasiparticle)
Hamiltonian $\tilde{H} = U H U^\dagger$ all matrix elements connecting different energy sectors
of the Hilbert space are transformed to zero to a given order in $1/\Delta$. The initial condition for $\tilde{H}$ is obtained by calculating $|\tilde \Psi(t=0)\rangle = U |\Psi(t=0)\rangle$.
Using that the term $-\frac{J}{2} \sum_i S^+_{0,i} S^-_{\pm 1,i}$  in $H$ raises the energy by $2 \alpha \Delta J$ producing the configuration shown in Fig.~\ref{fig1}b, we
obtain to leading order in $1/\Delta$
 \begin{eqnarray} \label{SWtrafo}
 |\tilde \Psi(t=0)\rangle  &=&  e^{S} |\Psi(t=0)\rangle  \\
S &\approx& -\frac{1}{4 \alpha \Delta}  \sum_j \left(S^+_{0,j}( S^-_{ 1,j}+ S^-_{- 1,j})-h.c.\right) \nonumber
\end{eqnarray}
In the quasiparticle picture, the initial wave function can therefore
be viewed as a dressed line of down-spins, with excitations of the type shown in Fig.~\ref{fig1}b.
Their density is 
\begin{equation} \label{rho}
\rho \approx \frac{2}{(4 \alpha \Delta)^2} \ll 1
\end{equation}
They can either be viewed as a kink-antikink pair located on neighboring sites or, equivalently, as a hole and a free spin again on neighboring 
sites. $\rho$ is therefore the density of pairs of quasiparticles.
 As $\rho \ll 1$ the probability that during the time evolution more than two quasiparticles come close to each other, is very small.
Therefore, most of the  quantum dynamic, at least on short length scales, can be tackled by solving quantum-mechanical 2-particle problems.
First, one has to consider the initial state problem.
Starting from the state  $S^+_{0,j} S^-_{ 1,j}|\Psi(t=0)\rangle$ (see Fig.~\ref{fig1}b) one solves the two-particle problem in the Hilbert space containing either one hole and one free spin ($h$+$f$)
or one kink and one antikink ($k$+$\bar k$).
\begin{eqnarray}
S^+_{0,j} S^-_{ 1,j}|\Psi(0)\rangle \to \left\{ \begin{array}{c} h+f \\k +\bar k\end{array} \right.
\end{eqnarray}
Details of the calculation are given in Appendix \ref{app_iniw}. The result is that for $\alpha=2$ in $89.8\%$ of all cases a $k\bar k$ pair is produced,
while only in $10.2\%$ a spin is emitted into the $2d$ continuum leaving a hole behind.
The resulting momentum distributions of kink-antikink pairs and holes in shown in Fig.~\ref{fig5} and discussed in section \ref{results}.
Due to the translational invariance of the initial state in $y$-direction, the momenta of a quasiparticle pair are opposite, for example $q$ and $-q$.

While an emitted down-spin never interacts with other quasiparticles, this is not the case for the time evolution
of $k$, $\bar{k}$ and $h$ which move along the line, thereby hitting other excitations with probability $1$.
We therefore have solved a series of two-particle scattering problems 
\begin{eqnarray}
k + \bar k &\to& \left\{ \begin{array}{c} k + \bar k   \\ h + f\end{array} \right., \quad 
k + h \to \left\{ \begin{array}{c} k + h   \\ \delta^{(1,2)} + f \end{array} \right., \nonumber \\
h + h &\to& \left\{ \begin{array}{c} h + h   \\ \delta^{(0,3)} + f \end{array} \right.,\quad
k + k \to k+k  
\end{eqnarray}
As more complex excitations, $\delta^{(n,m)}$, are produced in these initial reaction processes we need further scattering processes.
The scattering events involving more complicated excitations have an even richer structure with a
longer list of possible outgoing channels, see Appendix \ref{app_scatt} for an example.
For each of these processes we have calculated the transmission, reflection and reaction rates as function of the incoming and outgoing momenta.
For $|n|\geq3$ or $m>10$ we, however, neglect all transmission channels which become very small for such defects.
Note that a related calculation has recently been performed in a purely one-dimensional system by Ganahl, Haque and Evertz \cite{Ganahl13}.
We checked that we recover their result for $\alpha \to \infty$.
Due to the integrability of the xxz chain in $1d$ (broken for $\alpha < \infty$) one finds always perfect transmission in this limit.
Similar to \cite{Ganahl13}, we find that during a transmission event defects of the form $\delta^{m,n}$ are shifted by one (kink or antikink transmission) or two (hole transmission) lattice sites.
The numerical solution of all these two-particle problems provides a list of reaction and scattering rates. But it does not yet solve the strongly interacting many-particle
problem arising from the correlations induced by the scattering processes. 

\section{Semiclassical dynamics}

To solve the many-particle problem, we again use that the density of excitations is very small and their momenta are large (at least initially).
This implies, that the distance of defects will typically be much larger than their wavelength.
Therefore, the time evolution between scattering events can be described classically.
This type of argument, originally introduced by Damle and Sachdev \cite{Sachdev97} has been used successfully
in a series of papers calculating correlation functions, see, e.g., \cite{Sachdev97b,akos06}, and time evolution in integrable \cite{cardy07,Rieger11} and non-integrable \cite{tails} gapped one-dimensional systems.
In the low density limit considered here, only quasiparticles which are created in a pair are entangled.
As shown by Calabrese and Cardy \cite{cardy05}, this can be used to estimate the time dependence of the entanglement entropy.

For our semiclassical calculation we use the following algorithm: First, with probability density $\rho$, Eq.~(\ref{rho}), we randomly choose positions where initially pairs of excitations are created.
The particle type and the momenta of these pairs are randomly drawn from the quantum-mechanically calculated distribution function, see above.
These particles propagate with their respective group velocity $\partial_q \varepsilon^{k/h}_q$ with $\varepsilon^{h}_q \approx  -\alpha J \cos(q)$ and
$\varepsilon^{k/\bar k }_{q} \approx -J \cos(q) $. It is not necessary to keep track of the emitted free spins as they never scatter with other excitations.
When, however, two of the other quasiparticles hit each other, we randomly choose an outgoing channel according the quantum-mechanical reaction-, emission- and transmission rates and update particle type and particle momenta accordingly. 
An example is shown in Appendix \ref{app_scatt}.
A typical snapshot of such a time evolution is shown in Fig.~\ref{fig2}.

The numerical implementation is accomplished by keeping track of two lists, one encoding excitation types, their positions and momenta, the other the times of future scattering events
calculated from the positions and velocities of quasiparticles. In a typical numerical simulation we treat of the order of $10^5$ quasiparticles, time-evolving the system until only about 10 mobile defects are left.
While on average, the particles evaporate after 3 or 4 collisions, the slowest particles perform millions of collisions, see below. To improve statistics, we averaged over $800$ runs.

 \begin{figure}[t]
 \includegraphics[width=0.95 \linewidth]{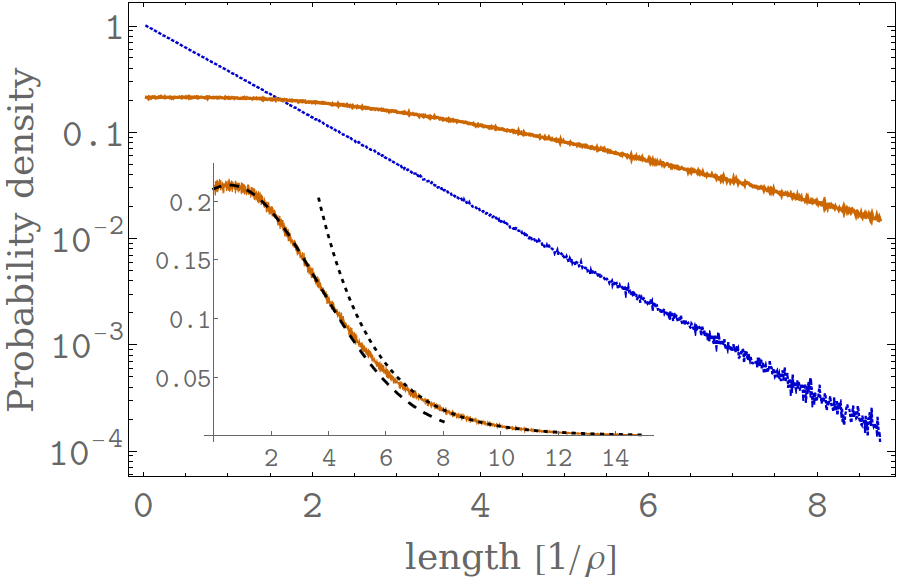}
 \caption{
 Probability distribution of the length of lines of down spins directly after the quench (dark blue line), and at the end of the simulation (light orange line), when there are no mobile particles left.
The initial distribution is exponential with decay length $1/\rho=32 \Delta^2$. 
The final distribution is exponential only for large ($l \gtrsim 8 /\rho $) line segments (dotted line in the inset)
and decays on the length scale $ (1.98 \pm 0.02)/\rho$.
A broad maximum is located at $ (0.51 \pm 0.05)/\rho$.
\label{fig3} }
 \end{figure}

\section{Results}\label{results}

In the following, all distances are measured in units of $r_0=1/\rho=8 ( \alpha \Delta)^2$, the initial average distance of pairs of defects, and an associated typical scattering time $ \tau_0 = r_0/J$.
Then, for $\Delta \gg 1$ all results are independent of $\Delta$. 
All numerical values do, however, depend on the spatial anisotropy $\alpha$, see Eq.~(\ref{ham}), and are given for $\alpha=2$.

Qualitative features of the time evolution after the quench can already be seen in Fig.~\ref{fig2}. By evaporation (yellow dots) of free spins, the number of mobile excitations
drops as function of time and mainly immobile defects (black worldlines) remain for longer times. 

We first investigate the properties of the system in the limit of large times, $t \to \infty$. In this limit, all mobile one-dimensional defects ($k$, $\bar k$, $h$) are gone.
They have emitted free down-spins and thereby contributed to the formation of larger, immobile defects. 
The total number $N_f$ of emitted free spins is given by $N_f \approx  1.7 \rho N_\downarrow$, where $N_\downarrow$ is the total number of down spins (i.e., the total length of the initial line).
A schematic picture  of the remaining system is shown in Fig.~\ref{fig1}e. It consists of line segments with an average length
of $3.03 /\rho \gg 1$ (for better visibility, length scales are {\em not} correctly shown in the schematic picture of Fig.~\ref{fig1}e). The end down-spins of these segments have an average spacing of $5.02$ lattice constants.
The distribution of the size of line segments is shown in Fig.~\ref{fig3} (the distribution of immobile defects
is discussed in Appendix \ref{app_immob}).  Fig.~\ref{fig3} can be viewed as a distribution function describing the many-particle bound  states (or quasi-bound states, see above), where on the $x$ axis the number of down spins contributing to each bound state is shown. 
The distribution function is very broad with an exponential tail.
In contrast, the initial distribution of distances of defects is purely exponential (blue line in Fig.~\ref{fig3}), as it is created by a Poisson process
(the number of quasiparticle pairs is Poisson-distributed, as can be seen by inspection of Eq.~(\ref{SWtrafo})).
 \begin{figure}[t]
 \includegraphics[width=0.95 \linewidth]{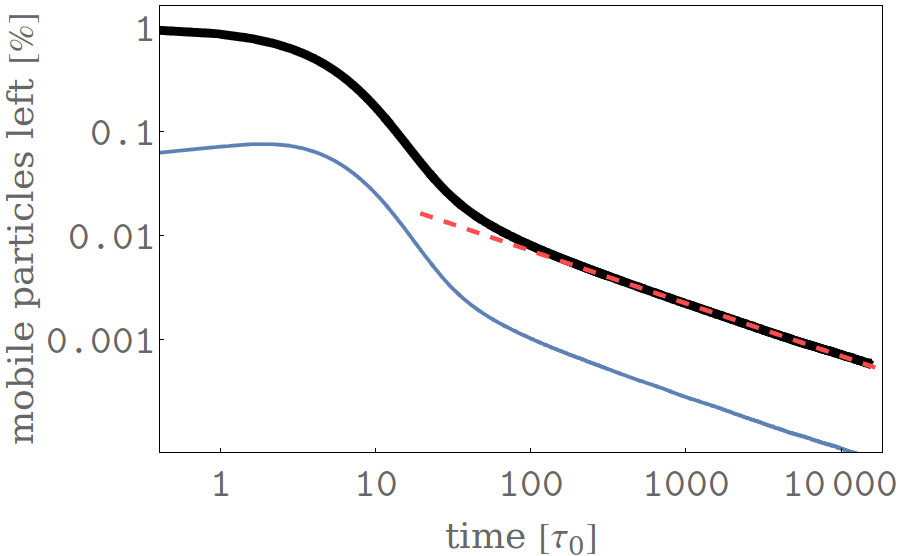}
 \caption{
Time evolution of the percentage of mobile particles remaining ($k$, $\bar k$ and $h$: black thick line). 
It decays exponentially for short times, while there is an algebraic long time tail $\sim (\tau_0/t)^{1/2} $ for long times $ t \gtrsim 50 \, \tau_0 $ (red dashed line).
The lower blue thin line shows which fraction of the mobile particles are holes. It increases for short times  $ t \lesssim 10 \, \tau_0 $ due to annihilation of kink-antikink
pairs, which produces mobile holes. 
\label{fig4} }
 \end{figure}

\begin{figure}[t]
\begin{center}
\includegraphics[width=0.95 \linewidth]{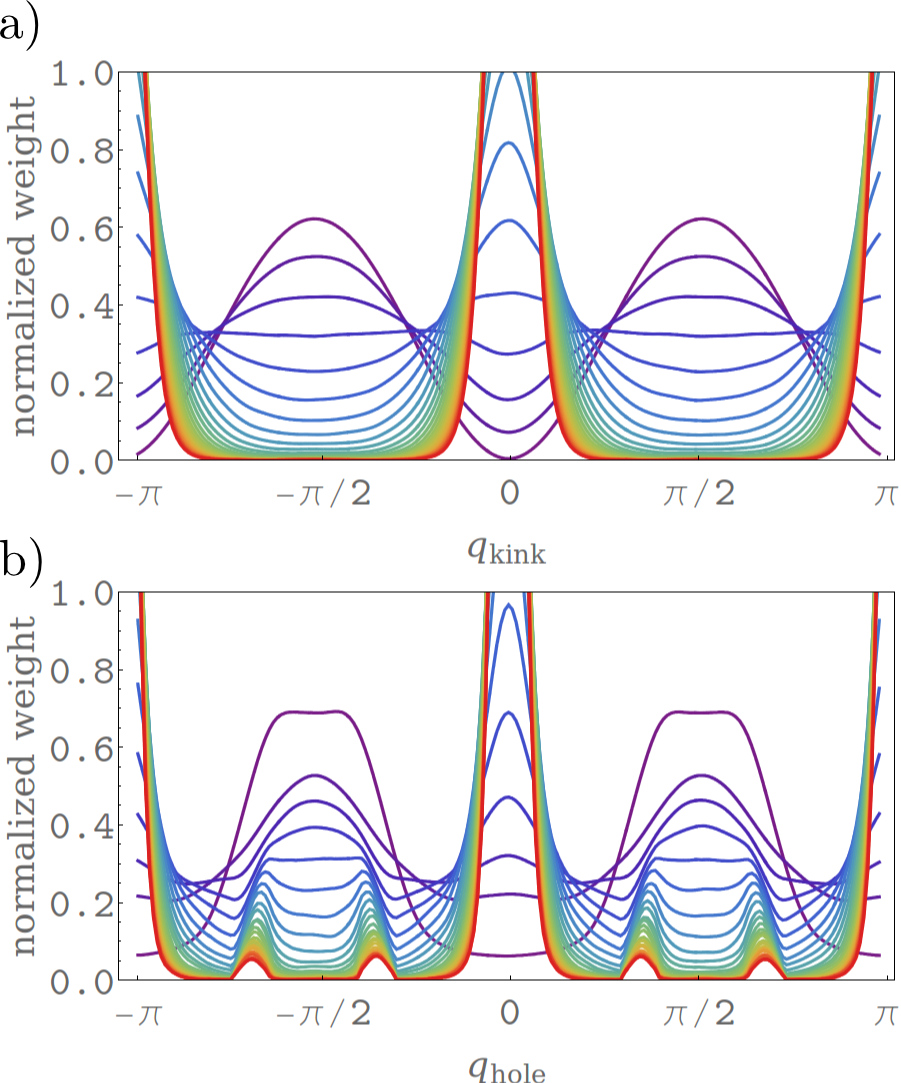}
\end{center}
 \caption{
Time evolution of the kink and antikink (panel a) and hole (panel b) momentum distributions from $t=0$ (purple lines) to $ t\approx 100 \,\tau_0 $ (red lines). For better visibility,  each momentum distribution is normalized, $\int n_q d q=1$. The 
 rapid drop of the total number of mobile particles is shown in Fig.~\ref{fig4}.
\label{fig5} }
 \end{figure}

The dynamics as a function of time is governed by three facts: (i) $kh$-, $\bar k h$- and $hh$-collision can lead to the emission of a free down-spin, $f$, in combination with the creation of larger immobile defects.
Those accumulate and grow in size by further emission processes. By this mechanism, the number of mobile defects decreases in time. 
(ii) Fast defects have a higher probability to hit other defects, and (iii) the probability to emit a free spin strongly depends strongly on momentum and is small when slow particles scatter, while the probability of reflexion is high, see Appendix \ref{app_scatt}.
This implies that the fast mobile particles evaporate most efficiently such that only slow defects with momenta close to $0$ and $\pm \pi$ remain.
To calculate the inverse lifetime $1/\tau_q$ of a mobile defect with small velocity $v_q \propto q$,
we note that the rate to hit another defect is of the order of $v_q/\rho \sim q$. 
Furthermore, the probability of emission turns out to be generically linear in $q$ for small $q$, such that
\begin{equation}
1/\tau_q \sim J  q^2 
\end{equation}
and likewise for $q \approx \pm \pi$.
Therefore, the time-dependence of the density of mobile defects is approximately given by
\begin{eqnarray}
n_{k,\bar k,h} \sim \int dq\,e^{-t/\tau_q} \sim 1/\sqrt{J t} \label{long}
\end{eqnarray}
consistent with our numerical simulation as can be seen in Fig.~\ref{fig4}. Here the number of mobile quasiparticles is shown  as function of time. 
After a period of evaporation by rapidly moving excitations, the long time tail described by Eq.~(\ref{long}) governs the physics for $t\gtrsim 50 \tau_0$.
This physics is also reflected in the momentum-distribution function of kinks and holes shown in Fig.~\ref{fig5}. 
The initial momentum distribution of kinks is governed by the fact that the $k\bar k$-pairs are initially created on neighboring sites with an approximate $\sin^2 q$ momentum distribution.
Scattering leads to a redistribution of momenta. As particles with fast velocities evaporate rapidly in collisions, peaks close to $0$ and $\pm \pi$ form in the momentum distribution.
In the regime,  $t\gtrsim 50 \tau_0$, where these peaks dominate the momentum distribution, we also observe the long-time tail in Fig~\ref{fig4}.
The peculiar initial momentum distribution of the holes, purple line in Fig.~\ref{fig5}b, with a pronounced plateau for momenta close to $\pi/2$ reflects, that their creation involves the
emission of a free down-spin into the $2d$ background, see Appendix \ref{app_iniw}. 
During the time evolution, small humps in the hole-momentum distribution, located to the left and right of $\pm \pi/2$, build up.
They arise from the creation of holes by the annihilation of slow kink-antikink pairs.

\section{Conclusion and outlook}

A main goal of this work was to study the statistical properties of many-particle bound states obtained after a quantum quench.
Experimentally, such properties can be probed with the newly developed quantum-microscopes \cite{greiner09, Weitenberg11}.
With such a microscope, one can rather directly measure the distribution of lines and clusters of spin-down states. Characterizing a quantum quench in this way,
contains much more information than can be obtained by considering averages of operators or 
correlation functions. 
The  formation of many-particle bound states is a question of relevance in many different areas of physics,
including the hadronization in a high-energy collision experiment or the question of how heavy elements are formed in supernova explosions.
While the physics is very different in those cases, one also has to rely on semiclassical and stochastic methods to obtain a theoretical description for such systems.

The semiclassical methods used in this paper are justified close to the Ising limit, $\Delta \gg 1$. More precisely, we have tracked the dynamics of the system for times of the order of 
\begin{equation}
\frac{1}{J} \ll t \ll \frac{\Delta^4}{J}
\end{equation}
Shorter time-scales describe the dynamics of the formation of quasiparticles, which we did not consider (but is easy to compute).
In the long time limit, however, several effects beyond the present approximation become important.

To begin with, the typical wavelength of the mobile quasiparticles ($k$, $\bar k$ and $h$) increases with $\sqrt{t}$ (as $1/\tau_k \propto k^2$) and
becomes of the order of the distance of defects, $\sim 1/\rho \propto \Delta^2$ at $t \sim \frac{\Delta^4}{J}$. This implies that the semiclassical approximation and the $1/\sqrt{t}$ law
obtained in Eq. (\ref{long}) ceases to be valid in this regime.

More importantly, for long times also the motion of larger defects ($\delta^{(0,3)}$, $\delta^{(\pm 1,2)}$, ...) and bound states with a small number of down-spins have to be considered.  
For example, in our calculation $\sim 15$\% of the defects are of type $\delta^{(0,3)}$ and $\sim 16$\%  are of type $\delta^{(\pm 1,2)}$, see Appendix \ref{app_immob}.
These two are the fastest defects, with hopping rates of order $J/\Delta^2$, which remain when practically all mobile defects with hopping rates of order $J$ have been evaporated. 
The heavier defects hit another defect after a typical time of order $\Delta^4/J \sim \tau_0 \Delta^2$. They can emit 
not only single free down-spins but also bound states of, e.g., two or three down-spins which evaporate into the background with hopping rates of order $J/\Delta$ or $J/\Delta^2$, respectively.
By this mechanism, a large number of these two-particle bound states (comparable to the number of single free spins) will be created, if sufficient long times are considered.

For large $\Delta$ these effects can safely be ignored on typical experimental time scales.
For smaller $\Delta>1$, however, one can indeed expect that after an initial burst of emitted single spins, 
two and three down-spin clusters can be observed which are not described by the distribution function calculated by us to leading order in $1/\Delta$. 

The experimental possibility to perform projective quantum measurements on many-particle systems opens new opportunities to characterize many-particle systems in or out of equilibrium.
It is a challenge for future work to develop theoretical methods to describe the statistics of measurements beyond the semiclassical limit considered in this paper.

\section*{Acknowledgments}
We thank I.~Bloch, M.~Haque, U.~Schneider and, especially, H.~G.~Evertz, for useful discussions and the DFG for financial support within CRC TR12.

\appendix

\section{Calculation of the initial momentum distributions} \label{app_iniw}

We consider the Hamiltonian Eq.~(\ref{ham}) in the limit $\Delta \gg 1 $, and the initial state shown in Fig.~\ref{fig1}a. As described in Sec.~\ref{zoology}, we introduce quasiparticles using a Schrieffer Wolff transformation, Eq.~(\ref{SWtrafo}). Within the quasiparticle language, the initial state can be viewed
as excitations of the type shown in Fig.~\ref{fig1}b, occurring with low density, Eq.~(\ref{rho}). From this initial states, kinks, antikinks, holes and free spins ($k$, $\bar k$, $h$, $f$) are created, whose momentum distribution will be calculated in the following.

As $\Delta \gg 1$, we consider the Hilbert space of a single pair of excitations  with fixed energy to leading order in $\Delta$. The energy of the initial state shown in Fig.~\ref{fig1}b is (relative to the state of Fig.~\ref{fig1}a) $2 \Delta \alpha J$. For $\Delta \to \infty$ it is degenerate to states of the type shown in Fig.~\ref{fig1}c and d which span the Hilbert space considered in the following.
As the initial state is translational invariant in y-direction and symmetric under the mirror transformation $x \to -x$, we have to consider the mirror-symmetric $ K_y = 0 $ ($ K_y $: total momentum in y-direction) sector of this Hilbert space only. The two-particle problem can be reduced to a single-particle problem by using
center-of-mass and relative coordinates. For notational convenience, we write the resulting single-particle problem in the language of second quantization where $\hat{a}^\dagger_q$ is the creation operator for the $k\bar k$ pair
in momentum space while $\hat{b}^\dagger_{\bm k}$ creates a $fh$ pair. Note that $q$ is one-dimensional while $\bm k$ is a two-component vector as the free spin is mobile in two dimensions. We use a convention where $0\le q,k_x \le \pi$ while $ -\pi \le k_y \le \pi$. $\hat{c}^\dagger$ creates the
initial configuration (the mirror-symmetric and translational invariant version of the state shown in Fig.~\ref{fig1}b).

The effective Hamiltonian to leading order in $ 1/\Delta $ is given by
\begin{widetext}
\begin{align}
 \hat{H}_e = \sum\limits_q \varepsilon^{k \bar k}_{q,K_y=0} \hat{a}^\dagger_q \hat{a}_q + \sum\limits_{k_x,k_y} \varepsilon^{hf}_{\bm k,K_y=0} \hat{b}^\dagger_{\bm k} \hat{b}_{\bm k}
 -J \sum\limits_q \sin(q) (\hat{a}^\dagger_q \hat{c} + \hat{c}^\dagger \hat{a}_q) - \frac{J}{2} \sum\limits_{k_x,k_y} \sin(k_x) (\hat{b}^\dagger_{\bm k} \hat{c} + \hat{c}^\dagger \hat{b}_{\bm k})
\end{align}
\end{widetext}
where 
\begin{align}
 \varepsilon^{k \bar k}_{q,K_y} & = -2 J \cos(\frac{K_y}{2}) \cos(q) \nonumber \\
\varepsilon^{hf}_{\bm k,K_y} & = -J \cos(k_x) - 2 \alpha J \cos(\frac{K_y}{2}) \cos(k_y)  \nonumber
\end{align}
are the kinetic energies of a kink-antikink pair ($k \bar k$) and a hole+free spin pair ($hf$), respectively.
We are interested in the momentum distribution of kinks and holes for times long compared to $1/J$.
Therefore we have to calculate
\begin{align}
P_{k\bar k} (q) & = \lim_{t \to \infty} \;| \langle 0 |  \hat{a}_q  |\Psi(t) \rangle|^2 \label{Wki} \\
P_{hf} (k_y) & = \int\limits_0^\pi \frac{d k_x}{\pi} \; \lim_{t \to \infty} \; | \langle 0 |  \hat{b}_{\bm k}  |\Psi(t) \rangle|^2  \label{Who} 
\end{align}
where $ |\Psi(0) \rangle=\hat{c}^\dagger |0\rangle$ and $ |\Psi(t) \rangle=e^{-i \hat{H} t} |\Psi(0)\rangle$. Note that we do not need the $k_x$ distribution of free spins, therefore we just sum over this variable.
\begin{figure}[thb]
\includegraphics[width=0.95 \linewidth]{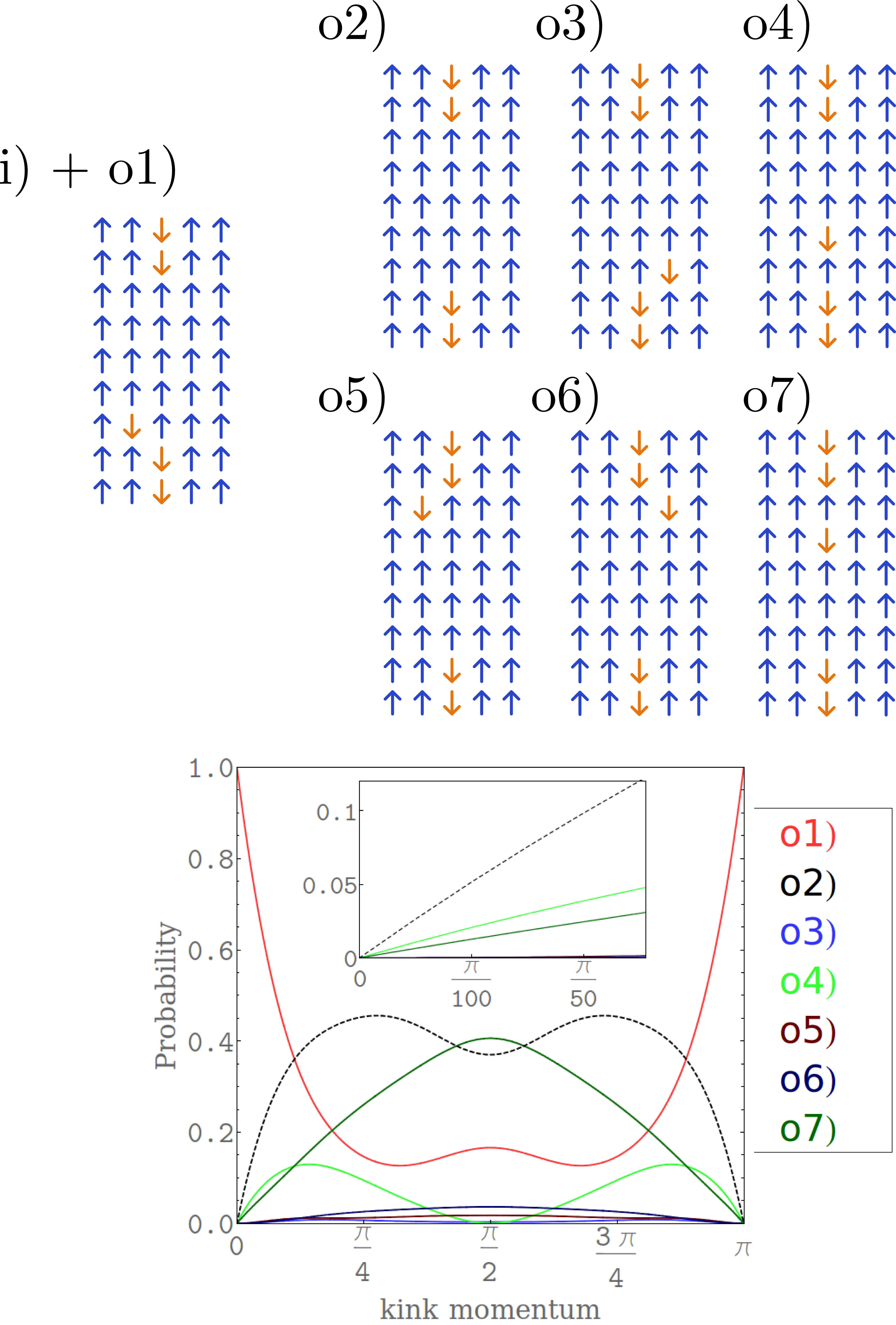}
 \caption{
Scattering rates of a kink from an immobile $\delta^{(1,4)}$ defect as a function of the incoming momentum of the kink. There are seven outgoing channels o1)-o7) sketched above the figure. For small momenta and momenta close to $\pi$ the reflection probability o1) approaches $1$. In the same limit, all other scattering  and emission rates vanish linearly with momentum as is shown in the inset.
 \label{fig_scattfull} } 
  \end{figure}
Using standard manipulations, we obtain
\begin{align}
 P_{k\bar k} (q) & = \frac{2 J^2 \; \sin^2 (q)}{|\varepsilon^{k \bar k}_q- J^2 \, G^{k \bar k}_0 (\varepsilon^{k \bar k}_{q,0},0) - \frac{J^2}{4}\, G^{hf}_0 (\varepsilon^{k \bar k}_{q,0},0)|^2} \nonumber \\
 P_{hf} (k_y) & = \int\limits_{0}^\pi \frac{d k_x}{\pi} \; \frac{\frac{1}{2} J^2 \; \sin^2 (k_x)}{|\varepsilon^{hf}_{\bm k,0}- J^2 \, G^{k \bar k}_0 (\varepsilon^{hf}_{\bm k,0},0) - \frac{J^2}{4}\, G^{hf}_0 (\varepsilon^{hf}_{\bm k,0},0)|^2} \nonumber 
\end{align}
where we used the  local Green functions
\begin{align}
G^{k \bar k}_0 (\omega,K_y) &= \int\limits_0^\pi \frac{d q}{\pi} \frac{2 \sin^2(q)}{\omega - \varepsilon^{k \bar k}_{q,K_y} + i \epsilon} \\
G^{hf}_0 (\omega, K_y) &= \int\limits_{-\pi}^\pi \frac{d k_y}{2 \pi} \int\limits_0^\pi \frac{d k_x}{\pi} \frac{2 \sin^2(k_x)}{\omega - \varepsilon^{hf}_{\bm k,K_y} + i \epsilon}   \label{gh}
\end{align}
We have checked numerically that  $ \int_0^\pi  \frac{d q}{\pi} P_{k\bar k} (q) + \int_{- \pi}^\pi   \frac{d k_y}{2 \pi} P_{hf} (k_y) = 1 $.

\section{Calculation of the scattering rates} \label{app_scatt}

We have calculated the two-particle scattering rates for more than $70$ settings. As in Appendix \ref{app_iniw}
we reformulate the two-particle scattering problem in c.~o.~m.~and relative coordinates and solve the appropriate two-dimensional scattering problem.  The scattering rates are obtained from the conservation of energy, lattice momentum and the probability current.
We do not consider any virtual states during the scattering, as these processes are of higher order in $1/\Delta$. As we are not interested in the momentum $k_x$ of the outgoing free spin, this degree of freedom can be integrated out and enters the scattering problem only via the Green function (\ref{gh}).

As an example, we give here the explicit formula for the momentum distribution, $P^{k\bar k \to fh}_{k_1,k_2}(q)$, for the creation of a hole with momentum $q$ 
in a scattering process of a $k\bar k$ pair with momenta $k_1$ and $k_2$:
\begin{widetext}
\begin{align} \label{scmom}
 P^{k\bar k \to fh}_{k_1,k_2}(q) = & \frac{8 J |\cos(\frac{k_1+k_2}{2}) \sin(\frac{k_1-k_2}{2})| \sqrt{J^2-(E_S(k_1, k_2)+2 \alpha J \cos(\frac{k_1+k_2}{2}) \cos(\frac{k_1+k_2}{2}-q))^2}}{|2 J \cos(\frac{k_1+k_2}{2}) \exp(i \frac{|k_1-k_2|}{2}) + \frac{J^2}{2} G_0^{hf} (E_S(k_1,k_2), k_1+k_2)|^2} \\ 
& \Theta \left( J-|E_S(k_1, k_2)+2 \alpha J \cos(\frac{k_1+k_2}{2}) \cos(\frac{k_1+k_2}{2}-q)| \right) \nonumber 
\end{align}
\end{widetext}
where $E_S(k_1, k_2)= -J (\cos (k_1) + \cos (k_2)) $ is the energy of the incoming $k \bar k$ pair, $\Theta$ denotes the Heaviside step function,
and $\int\nolimits_{-\pi}^\pi \frac{dq}{2 \pi} P^{k\bar k \to fh}_{k_1,k_2}(q) $ is the total emission probability.
The function in Eq.~(\ref{scmom}) depends on three momenta.
Fortunately, in all other non-trivial scattering processes at least one immobile particle is involved and they can therefore be described by functions of one or two momenta which can be tabulated.
An example is given below. For large defects $\delta^{(n,m)}$ ($|n| \ge  3$ or $m > 10$) only the reflection and emission channels have been taken into account.

In Fig.~\ref{fig_scattfull} an example with $7$ outgoing channels is shown. 
We consider a kink scattering with an immobile $\delta^{(1,4)}$ defect.
The outgoing channels are shown in the upper part of Fig.~\ref{fig_scattfull}, and their associated "reaction schemes" are:
\begin{align} \nonumber \label{scattchannels}
 k + \delta^{(1,4)} \to
\left\{ \begin{array}{lr}  
 k+\delta^{(1,4)} & \text{(o1)}  \\
 f+ \delta^{(0,5)} & \text{(o2)} \\ 
 \bar{k}+ \delta^{(-1,4)} & \text{(o3)} \\
 h+ \delta^{(0,3)} & \text{(o4)} \\
 \delta^{(-1,4)}+\bar k & \text{(o5)} \\
 \delta^{(1,4)} + k  & \text{(o6)} \\
 \delta^{(0,3)}+h  & \text{(o7)}
\end{array}  \right.
\end{align}

In channels o1, o3 and o4 a particle is reflected, o2 describes an emission event, while o5, o6 and o7 are situations where a mobile particle is transmitted. 
Both during reflection and transmission the type of particle can change.
As there is only one mobile particle in each of the incoming and outgoing channels, the momentum of the outgoing particle is fixed by energy conservation, except for channel o2.
Here, the emitted free down-spin has two degrees of freedom. As in the latter case, we do not need to know the outgoing momenta, but we are only interested in the total emission rate. 
All reflection, transmission and emission rates can be computed numerically as function of the incoming momentum of the kink.
These functions are computed only once and are tabulated to be used in the semiclassical simulations.
For the example given above, they are shown in the lower part of Fig.~\ref{fig_scattfull}. For fast incoming kinks ($ |q| \approx \pi/2 $) emission and transmission channels dominate, 
while for slow incoming kinks ($ q \approx 0 $ or $ |q| \approx \pi $) the emission probability (generically) decreases linearly to zero (see inset).
Therefore, slow particles scatter very often before they are emitted. This leads to the algebraic long time tails, as discussed in the main text.

\section{Statistics of the immobile defects} \label{app_immob}

\begin{figure}[thb]
\includegraphics[width=0.9 \linewidth]{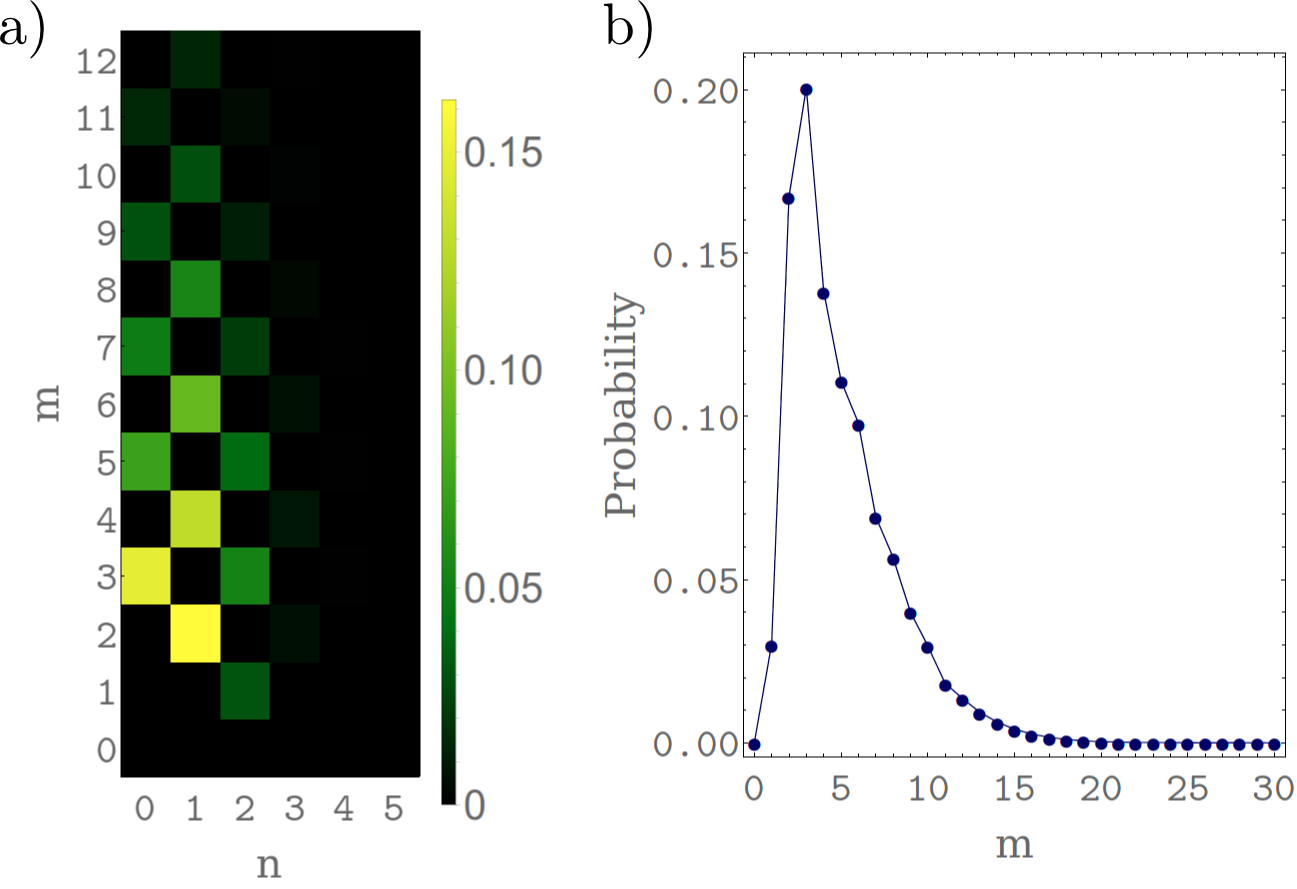}
 \caption{
a) Distribution $P_{\delta^{(n,m)}}$ of the immobile defects $\delta^{(n,m)}$ at the end of the simulation, when there are no mobile particles left.
$n$ and $m$ are measured in units of the lattice constant.
 c) The probabilities shown in b) summed over $n$.
 \label{fig_defstats} }
  \end{figure}

At the end of the simulation, when all mobile quasiparticles have been emitted, the $1d$ dynamics is frozen out.
In Fig.~\ref{fig1}e a schematic picture of a possible final configuration is shown (note that the length scales are not correct). 
The initial line is fragmented into pieces of many-particle (quasi- or true) bound states (orange down-spins). Separated from those, there are single down-spins
which have been emitted into the ferromagnetic background. The system can be characterized by the distribution function of down-spins participating to a bound state. This is shown in Fig.~\ref{fig3} of the main text.
On timescales where the bound states do not move, and the quasi-bound states do not decay, the system also has characteristic distribution of immobile defects, $\delta^{(n,m)}$, three examples are marked in Fig.~\ref{fig1}e.
The corresponding probability distribution $ P_{\delta^{ (n,m)}}$ is shown in Fig.~\ref{fig_defstats}a.
We find that only defects with odd $n+m$ are created (in an emission event, an incoming hole increases $m$ by $2$, while an incoming kink or antikink changes
both $n$ and $m$ by $1$, also all conversion processes follow this rule). This leads to the checkerboard pattern in the distribution in Fig.~\ref{fig_defstats}a.
In Fig.~\ref{fig_defstats}b the probability distribution, $\bar{P}(m) =\sum_{n}  P_{\delta^{ (n,m)}}$, of the gaps in $y$ direction, parallel to the line are shown. The expectation value is
$ \sum_{m} m \bar{P}(m)  \approx 5.02 $. 

\bibliography{bibq}

\end{document}